\documentclass[prd,preprint]{revtex4}
\usepackage[english]{babel}
\usepackage{amscd}
\usepackage{epsfig}
\usepackage{tabularx}
\usepackage{graphicx}
\usepackage{latexsym}
\usepackage{amsmath}
\usepackage{amsfonts}
\usepackage{amssymb}
\usepackage{orcidlink}
\usepackage{indentfirst}
\usepackage{hyperref}
\usepackage{times}
\usepackage[T1]{fontenc}
\usepackage{latexsym}
\usepackage{graphics}
\usepackage{verbatim}
\usepackage[absolute]{textpos}
\usepackage{wrapfig}
\usepackage{amsthm}
\usepackage{setspace}
\usepackage{color}
\definecolor{Red}{rgb}{0.9,0.1,0.1}
\definecolor{blue}{rgb}{0.25,0.25,0.6}

\newcommand{\be}{\begin{equation}}
	\newcommand{\ee}{ \end{equation}}
\newcommand{\ben}{\begin{eqnarray}}
	\newcommand{\een}{\end{eqnarray}}

\begin{document}
	
\title{Physics in the Public Square: University Extension as a Strategy for Integrating Physics Education and Science Communication}

\author{André A. A. Marinho$^{1,*}$\orcidlink{0000-0002-2570-3277}
	, Gisele B. Freitas$^1$\orcidlink{0000-0002-5729-3620} 
	and Camila B.C. da Silva$^1$ }

\affiliation{$^1$Centro de Ciências Exatas, Naturais e Tecnológicas - CCENT, Universidade Estadual da Região Tocantina
	do Maranhão (UEMASUL), R. Godofredo Viana 1300, 65901-480, Imperatriz, MA, Brazil}



\email{andre.afonso@uemasul.edu.br}

\begin{abstract}
	University extension activities play a fundamental role in bridging the gap between academia and society by fostering the socialization of scientific knowledge. This study reports and analyzes an outreach activity conducted in a public space, involving undergraduate students enrolled in Physics I, Physics III, and Physics IV courses within the Physics Teacher Education Program at the State University of the Tocantina Region of Maranhão (UEMASUL). The activity was developed through the design and presentation of didactic experiments using low-cost materials. Its main objectives were to disseminate fundamental physics concepts to the community, stimulate public interest in science, and provide pre-service teachers with a formative experience integrating theory, practice, and social responsibility. Data were collected from questionnaires adminisvelopment of communication skills, and the strengthening of the university’s social role, while also fostering scientific curitered to visitors $(n = 52)$. The results indicate that the activity significantly contributed to student learning, the deosity among participants.
	
\end{abstract}
\keywords{university extension; physics education; science communication; low-cost experiments; non-formal learning environments. }
\maketitle
\section{Introduction}

Physics is historically perceived by high school students and the general public as an abstract science, characterized by mathematically dense formulations that are often difficult to understand. This perception contrasts with the ubiquity of physical phenomena in everyday life, spanning domains from nanotechnology to astronomy, and from economics to pharmaceuticals \cite{lop, mor, joa}. This gap is further exacerbated by the scarcity of experimental activities in basic education, resulting from the lack of adequate laboratory infrastructure, limited opportunities for in-service teacher training, and insufficient access to specialized equipment \cite{adr}.

In this context, the curricularization of extension—established by Resolution CNE/CES No. 7/2018 of the Brazilian Ministry of Education \cite{bra} - mandates that undergraduate programs allocate at least 10\% of their total workload to extension activities integrated with teaching and research. At the State University of the Tocantina Region of Maranhão (UEMASUL), this commitment has been institutionalized through Resolution CONSUN No. 216/2022  \cite{uem}, which recognizes extension as a formative component for fostering socially engaged, critical, and context-aware professional practice, as discussed by Freitas and Pereira \cite{gis}.

There is a global concern to develop strategies in science education that promote cognitive learning \cite{pac}. Changes are needed in the teaching and evaluation of teacher performance in order to identify and correct shortcomings \cite{xli, dod}, that are expected and necessary in teaching and learning processes that deviate from the usual. Conducting activities in informal spaces \cite{bel,iza} and to seek different ways to strengthen the teaching of Physics and interdisciplinarity \cite{man}. An active approach as presented in the Investigative Science Learning Environment - ISLE method \cite{etk}. Like other proposals found in the literature, ours seeks the same ultimate goal: to take Physics beyond the walls of Universities and present to society the beauty of the science that describes natural phenomena, and the importance it has in our society with existing technologies and those still being developed.

Motivated by this institutional and pedagogical framework, the activity “Physics in the Public Square” was conceived as an outreach initiative in a non-formal public setting, based on the use of low-cost didactic experiments designed and implemented by pre-service physics teachers enrolled in the Physics Teacher Education Program at UEMASUL, Imperatriz campus. The initiative aimed to return university-generated knowledge to society by connecting academic knowledge with the lived reality of the local community.

This paper aims to report and analyze this activity, describing its planning and implementation processes, as well as the outcomes achieved, both in terms of student learning and its impact on the participating public.

\section{Theoretical Framework}

From a Freirean perspective, university extension is not limited to the unidirectional transfer of knowledge from academia to society; rather, it is understood as a dialogical process of collective knowledge construction \cite{fre}. Within this framework, pre-service teachers do not merely apply previously learned content, but actively learn and reinterpret their knowledge through engagement with diverse social contexts and actors.

Lopes \cite{lop} discusses the interrelationships among school, scientific, and everyday knowledge, emphasizing the need to expand science education beyond the physical boundaries of the classroom. When schools and universities engage with the broader community, they foster an interdisciplinary understanding of Physics, making it more meaningful and relevant to students and their families.

Non-formal educational spaces-such as public squares, parks, museums, and other community settings-are characterized by voluntary participation, the absence of curricular constraints, and the heterogeneity of audiences. These features require mediators to develop specific communicative competencies and to employ accessible language \cite{jac}. Such environments are particularly conducive to science communication, as they bring scientific knowledge closer to everyday experiences.

From this standpoint, the process of explaining occurs within language and cannot be reduced to the mere transmission of information from sender to receiver. Drawing on the reflections of Humberto Maturana \cite{mat}, science education can be understood as a dialogical process in which explanations are collectively constructed and socially validated. This perspective is especially relevant in outreach and science communication initiatives such as “Physics in the Public Square”, where undergraduate students assume the role of mediators of scientific knowledge for non-specialist audiences. In these contexts, explaining Physics concepts entails translating complex ideas into accessible forms, adapting language to diverse audiences while maintaining the epistemic criteria that underpin scientific explanations. The acceptance of an explanation thus emerges within a process of meaning negotiation, in which different experiences and perspectives are acknowledged — what the author \cite{mat} refers to as the “explanatory path of objectivity in parentheses.”

Furthermore, as highlighted by \cite{mat}, human actions are deeply intertwined with emotions, and engagement plays a crucial role in enabling shifts in the understanding of phenomena. Participation in science communication practices encourages students to connect Physics concepts with the everyday experiences of their audiences, thereby reshaping their own conceptual frameworks and enhancing their scientific communication skills. This process contributes not only to narrowing the gap between science and society, but also to the development of graduates capable of explaining, dialoguing, and adapting scientific language to diverse sociocultural contexts.

The use of low-cost experiments is widely recognized as an effective strategy in Physics education, both at the basic and higher education levels. According to \cite{adr}, such resources foster creativity, autonomy, and critical thinking, while enabling complex concepts to be demonstrated in concrete and visual ways. Furthermore, the construction of these experiments by students themselves promotes meaningful engagement with theoretical content, aligning with Ausubel’s theory of meaningful learning, as revisited by Moreira \cite{mor}.

\section{Methodology}

This outreach activity was planned and implemented by undergraduate students enrolled in the Physics Teacher Education Program at UEMASUL, under the supervision of instructors from the Physics I and Physics IV courses. Students from Physics III voluntarily joined the initiative, broadening disciplinary representation and strengthening curricular integration.

The planning phase took place over several weeks prior to the event and involved periodic meetings to define the experiments, assign group responsibilities, construct the artifacts, and rehearse presentations. Students were organized into groups and instructed to design and build didactic experiments using low-cost materials, such as PET bottles, wires, mirrors, magnets, cardboard, cans, and balloons, among others. Each group was also responsible for developing accessible explanations tailored to a lay audience, which were tested and refined during peer rehearsal sessions.

\begin{figure}[htb]
	\centerline{
		\includegraphics[{angle=90,height=10.0cm,angle=270,width=10.0cm}]{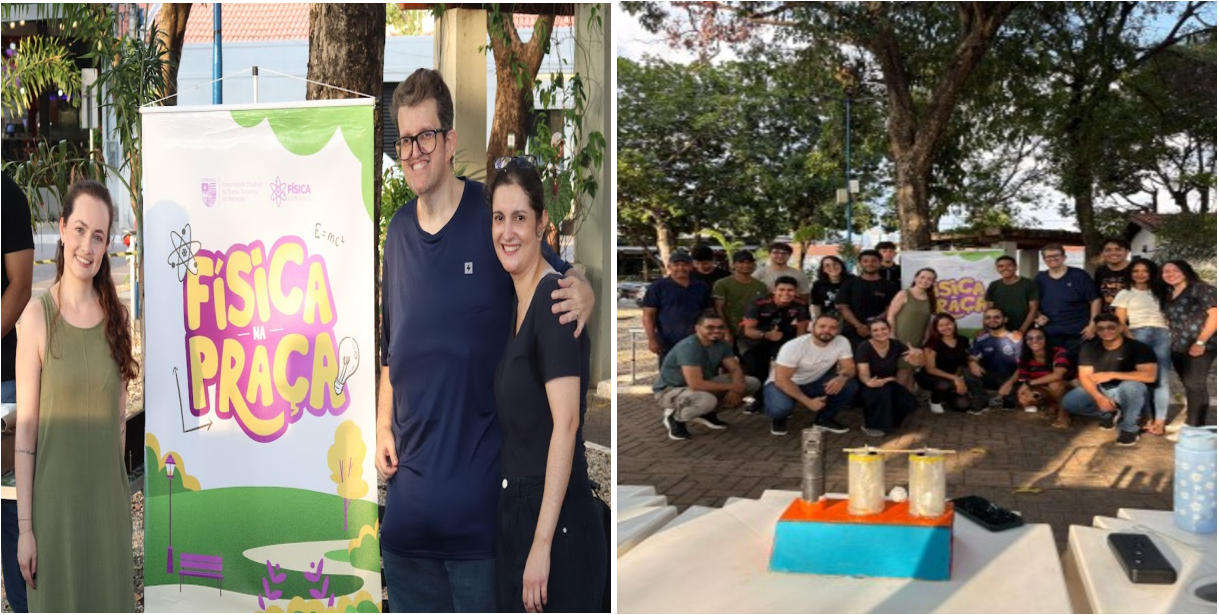}
	}
	\caption{{The professors who conceived the idea and our logo, \textbf{(left)}. Our team \textbf{(right)}.}}
	\label{fig1}
\end{figure}

The activity was held at Praça da Cultura, in the city of Imperatriz, Maranhão - Brazil — see (\ref{fig1}), on November 22, 2025, from 4:00 PM to 6:00 PM. The experiments were arranged in stations with tables and tents, allowing visitors to circulate freely. In total, 12 experiments were developed and exhibited, covering topics in classical mechanics (Physics I), electromagnetism (Physics III), and optics and electromagnetic waves (Physics IV). The event was publicized by UEMASUL’s Communication Office (ASCOM) and featured by two local television channels.

At the end of the activity, a structured questionnaire (\ref{app}) was administered to visitors, consisting of six closed-ended and two open-ended questions. Responses were tabulated and analyzed using a descriptive approach, with results presented in the Results and Discussion section. A total of 52 visitors voluntarily participated in the survey. Qualitative data from the open-ended questions were organized into emergent thematic categories.

\begin{figure}[htb]
	\centerline{
		\includegraphics[{angle=90,height=10.0cm,angle=270,width=10.0cm}]{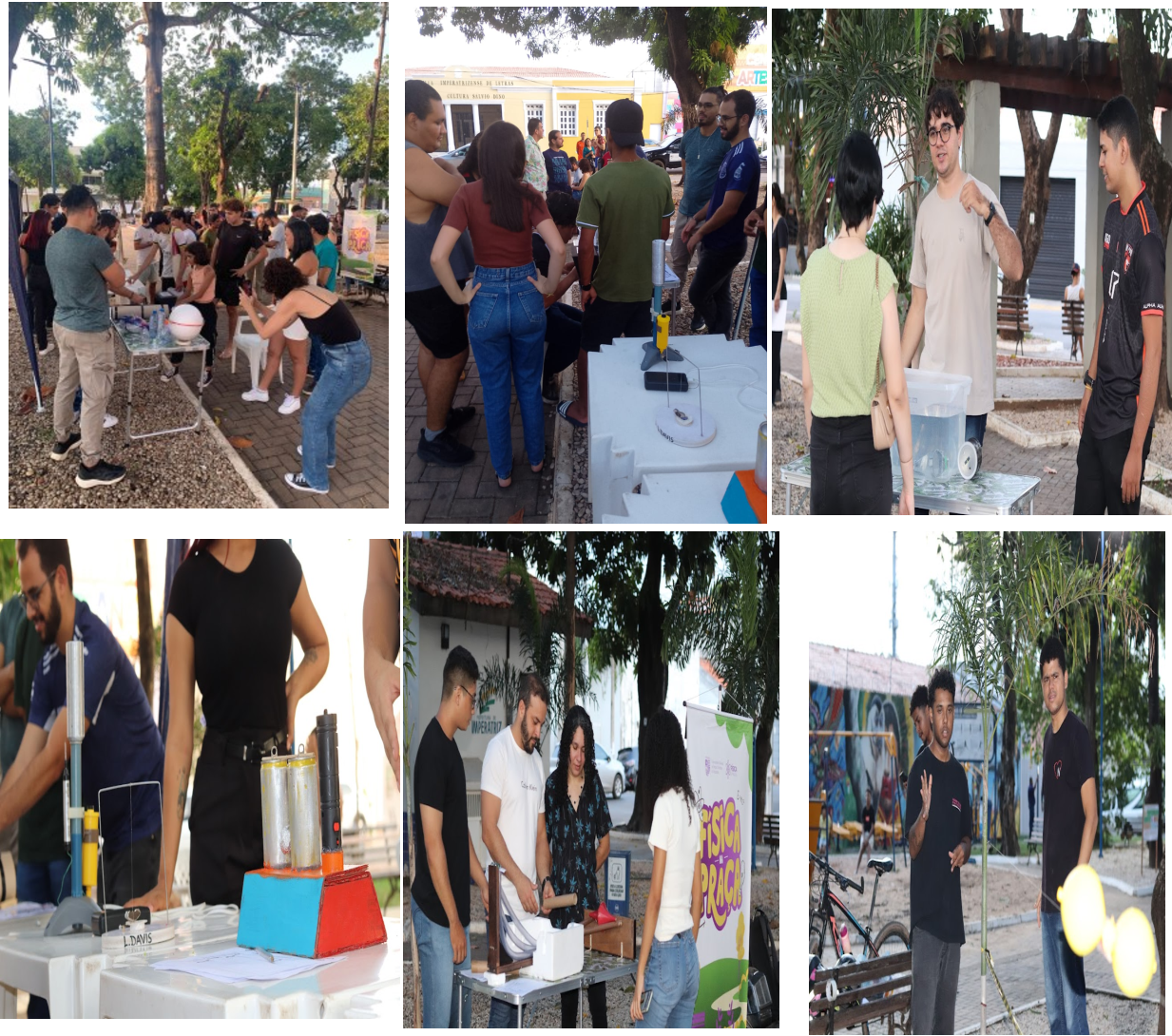}
	}
	\caption{{Some experiments presented by our team to the public.}}
	\label{fig2}
\end{figure}

The choice of a non-formal public space, rather than a partner school, was intentional. The goal was to engage a heterogeneous audience with no prior connection to the content, thereby posing an additional challenge for the pre-service teachers in terms of language adaptation and the management of unpredictable interactions. This decision was grounded in the understanding that teacher education is enriched when future educators are exposed to real and diverse contexts — see (\ref{fig2}).

\section{Results and Discussion}

During the activity, significant community engagement was observed, with participants including children, adolescents, and adults. The experiments stimulated curiosity and facilitated the understanding of physical concepts, highlighting the effectiveness of an experimental approach in a public, non-formal setting. The data collected through the questionnaire (n = 52) are presented and discussed below.

\subsection{Questionnaire Results}

The results for each question were as follows:

\begin{enumerate}
	\item Were you familiar with any of the experiments presented?
	Yes: 41 (78.1\%); No: 11 (21.9\%)
	
	\item Did the explanations help you understand the phenomenon?
	Yes: 50 (96.9\%); Partially: 2 (3.1\%); No: 0.
	
	\item Clarity of the explanations:
	Excellent: 42 (81.3\%); Good: 10 (18.8\%); Fair: 0; Poor: 0.
	
	\item What did you think of the hands-on experience?
	Very interesting: 50 (96.9\%); Interesting: 2 (3.1\%); Slightly interesting: 0; Did not like it: 0.
	
	\item Would you like more events like this?
	Yes: 52 (100\%); No: 0.
\end{enumerate}
.

\subsection{Open-Ended Responses}

Responses to the question “What did you learn today?” were qualitatively analyzed and organized into three categories: (i) specific Physics concepts; (ii) the relationship between Physics and everyday life; and (iii) scientific methodology.

In the first category, participants mentioned topics such as Archimedes’ principle/flotation (“how a submarine works”), Newton’s laws, electrostatics, optics, and density. In the second, participants demonstrated an increased awareness of Physics as part of daily life, as illustrated by statements such as: “[…] Physics in practice shows how much our reality is connected to science.” In the third category, responses highlighted the experimental nature of scientific knowledge, for example: “[…] it is possible to observe the effects of Physics using simple materials.”

Visitor suggestions emphasized the importance of continuing and expanding the initiative, including the incorporation of digital tools and the organization of similar events in other public spaces, targeting a broader range of age groups.

\subsection{Discussion}

The high approval rates—96.9\% indicating that the explanations facilitated understanding and 81.3\% rating their clarity as “Excellent”—suggest that the pre-service teachers were successful in communicating scientific concepts to a heterogeneous audience. These findings are consistent with Moreira’s \cite{mor} perspective on meaningful learning: by anchoring new concepts in concrete and visually accessible situations, the experiments supported effective cognitive assimilation among participants.

From the perspective of teacher education, the activity mobilized key competencies outlined in the National Curriculum Guidelines for teacher education programs, such as the ability to integrate theory and practice, adapt language to the target audience, and engage in collaborative work. The development of experiments using low-cost materials also fostered creativity and problem-solving skills—essential attributes for teaching in resource-constrained educational contexts, which are common in public school systems.

The integration of Physics I, Physics III, and Physics IV within a single initiative challenged the traditional fragmentation of disciplinary structures, promoting a more holistic and contextualized understanding of Physics as a scientific field. This approach aligns with the principle of interdisciplinarity advocated in the National Guidelines for Extension in Higher Education \cite{bra} and with the objectives of extension curricularization.

Finally, the interest and enthusiasm observed among participants—particularly children—and the unanimous demand (100\%) for more events of this nature reinforce the potential of university extension in non-formal settings to democratize access to scientific knowledge and to foster a scientific culture within the community.

\section{Conclusion}

The outreach activity “Physics in the Public Square” proved to be an effective strategy for integrating Physics Education with science communication in a non-formal public setting. The design and presentation of low-cost experiments by pre-service teachers enabled the consolidation of theoretical knowledge, the development of socio-educational competencies, and a meaningful rapprochement between the university and the community.

The collected data indicate a high level of satisfaction and learning among both students and visitors, reinforcing the relevance of extension activities as instruments for comprehensive education and for strengthening the social role of the university. The initiative also demonstrated that Physics can be communicated in an accessible and engaging manner to diverse audiences, contributing to its demystification.

It is recommended that similar initiatives be expanded and sustained, incorporating digital tools and participatory methodologies that address the needs of diverse audiences, including individuals with disabilities, in line with the suggestions provided by participants. The experience reported here may serve as a reference for other teacher education programs seeking to implement the curricularization of extension in innovative and socially committed ways.


{Acknowledgments}

The authors would like to thank the UEMASUL Communication Office (ASCOM) for their support in promoting the event, the local television channels for their media coverage, and the community of Imperatriz for their engagement and participation.


\appendix
\section{Appendix: Questionnaire Administered to Visitors}
\label{app}

\begin{enumerate}
	
	\item Were you familiar with any of the experiments presented? 
	
	( ) Yes      ( ) No

	\item	Did the explanations help you understand the phenomenon? 
	
	( )Yes   ( )Partially  ( )No

	\item How would you rate the clarity of the explanations?
	
	( ) Excellent   ( )Good  ( )Fair  ( )Poor
	
	\item What did you think of the hands-on experience?
	
	( )Very interesting   ( )Interesting  ( )Slightly interesting   ( )Did not like it
	
	\item Would you like more events like this?

	( )Yes   ( )No
	
	\item What did you learn today? 
	
	\item Suggestions:
	
\end{enumerate}

\end{document}